\documentclass[twocolumn,superscriptaddress]{revtex4-1}

\usepackage{graphicx}
\usepackage{amsmath}
\usepackage{amssymb}
\usepackage{setspace}
\usepackage{enumerate}
\usepackage{bm}
\usepackage{array}
\newcolumntype{C}[1]{>{\centering\let\newline\\\arraybackslash\hspace{0pt}}m{#1}}
\usepackage{microtype}
\usepackage{hyperref,color}
\usepackage{changes}

\newtheorem{theorem}{Theorem}

\usepackage{float}
\usepackage{amsbsy}
\usepackage{xcolor, ulem,url}

\begin{document}
\title{Control of connectivity and rigidity in prismatic assemblies}
\author{Gary P. T. Choi}
\affiliation{John A. Paulson School of Engineering and Applied Sciences, Harvard University, Cambridge, MA, USA}
\author{Siheng Chen}
\affiliation{John A. Paulson School of Engineering and Applied Sciences, Harvard University, Cambridge, MA, USA}
\author{L. Mahadevan}
\affiliation{John A. Paulson School of Engineering and Applied Sciences, Harvard University, Cambridge, MA, USA}
\affiliation{Department of Physics, Harvard University, Cambridge, MA, USA}
\affiliation{Department of Organismic and Evolutionary Biology, Harvard University, Cambridge, MA, USA}

\date{\today}

\newcommand{\xx}{\mathbf{x}}

\begin{abstract}
How can we manipulate the topological connectivity of a three-dimensional prismatic assembly to control the number of internal degrees of freedom and the number of connected components in it? To answer this question in a deterministic setting, we use ideas from elementary number theory to provide a hierarchical deterministic protocol for the control of rigidity and connectivity. We then show that is possible to also use a stochastic protocol to achieve the same results via a percolation transition. Together, these approaches provide scale-independent algorithms for the cutting or gluing of three-dimensional prismatic assemblies to control their overall connectivity and rigidity.
\end{abstract} 

\maketitle

Given a three-dimensional (3D) solid, how can we introduce cuts in it that convert it to a prismatic assembly that is either partially or fully connected, and can be either partially or completely rigid? Said differently, how does the topology of the underlying network of connectivity in such an assembly control the degrees of freedom (DoF) and the number of connected components (NCC)? And how can we use either deterministic or stochastic approaches to control both these properties? Here we explore and answer these questions using a combination of analysis and computation. In addition to being of intrinsic interest, the questions are of technological relevance for understanding the assembly of polyhedral building blocks into ordered structures in atomic systems~\cite{damasceno2012predictive} as well as the design of molecular materials~\cite{fujita2016self} and nanocrystals~\cite{henzie2012self}. The deterministic aspect of our study is related to the classical subject of rigidity theory~\cite{dehn1916starrheit} and self-assembly~\cite{philp1996self,sun2010self,RevModPhys.89.031001}. Using ideas from number theory allow us to provide algorithms for the control of rigidity and connectivity of prismatic structures in a hierarchical manner with direct consequences for structural assemblies. The stochastic aspect of our study is naturally related to bond and rigidity percolation~\cite{jacobs1995generic,garboczi1995geometrical,chubynsky2006self,briere2007self}. Complementing prior work on 3D rigidity analysis of networks~\cite{chubynsky2007algorithms}, our study of structural assemblies allows for the determination of the total and internal rotational DoF as well as the number and size of connected components, showing the existence of percolation transitions associated with the onset of connectivity and rigidity.

\begin{figure}[t]
\centering
\includegraphics[width=0.35\textwidth]{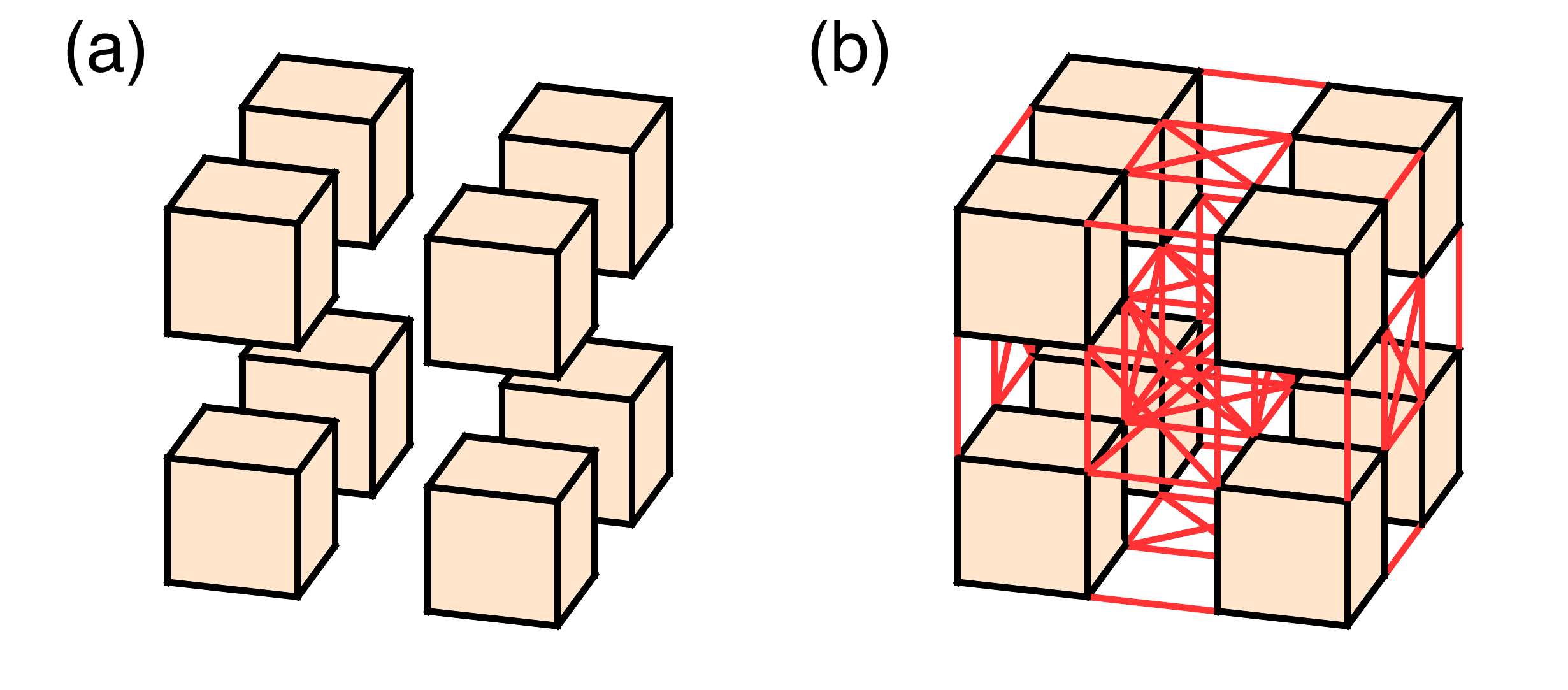}
\caption{{\bf Topological control of prismatic assemblies is analogous to controlling a linkage.} (a)~An example of how a solid can be decomposed into an $L\times M\times N$ 3D prismatic assembly (here $L = M = N = 2$). (b)~All possible links for connecting neighboring cubes in a $2\times 2 \times 2$ prismatic assembly. A subset of these can be used to control the number of connected components, while another can be used to control the number of internal degrees of freedom in the assembly.}
\label{fig:F1}
\end{figure}

To simplify our discussion, we start with a rectangular solid $D$ in $\mathbb{R}^3$ with parallel cuts introduced along equally-spaced grid lines in the $x$-, $y$- and $z$-directions. Assuming that the length, width and height of $D$ are all integer multiples of a positive number $l$, the cuts decompose $D$ into $L \times M \times N$ identical solid cubes with side length $l$ (Fig.~\ref{fig:F1}(a)). Then, we consider placing a number of infinitesimal links either deterministically or stochastically to connect some of the cubes with their neighbors (Fig.~\ref{fig:F1}(b)), thereby forming a 3D solid assembly. The infinitesimal links control the topology of the assembly and hence affect its rigidity and connectivity in terms of DoF and NCC. we can transform the cutting problem into a linkage problem which is similar in spirit to the setup in the planar analog - a kirigami structure~\cite{chen2020deterministic}.

 %%%%%%%%%%%%%%%%%%%%%%%%%%%%%%%%%%%%%%%%%%%
%\section{Deterministic control of connectivity and rigidity of a 3D cubic prismatic assembly}
We first explore the deterministic control of prismatic assemblies, and establish algorithmic protocols for determining the minimum number of links that can (i)~rigidify a prismatic assembly so that it has no internal modes of motion (i.e. control the DoF) or (ii)~connect a prismatic assembly (i.e. control the NCC).

%\subsection{Minimum rigidifying link patterns (MRPs)}
%\subsubsection{Calculating the DoF in a prismatic assembly}
Since each cube has three translational DoF and three rotational DoF, the maximum total DoF of any $L \times M \times N$ prismatic assembly is $d = 6LMN$. If all links are added, the entire prismatic assembly is rigid and hence the minimum DoF is $d = 6$. 

By the Dehn's rigidity theorem~\cite{dehn1916starrheit}, any closed convex polyhedron with infinitesimally rigid faces is infinitesimally rigid. Therefore, for each solid cube with side length $l$, there are exactly twelve \textit{edge length constraints} in the form of
\begin{equation}
g_{\text{edge}}({\bm v}_i,{\bm v}_j) = \|{\bm v}_i -{\bm v}_j\|^2 - l^2 = 0,
\end{equation}
where ${\bm v}_i$ and ${\bm v}_j$ are two adjacent vertices in a cube, and six \textit{diagonal length constraints} for all faces of the cube:
\begin{equation}
g_{\text{diagonal}}({\bm v}_i,{\bm v}_j) = \|{\bm v}_i -{\bm v}_j\|^2 - 2l^2 = 0,
\end{equation}
where ${\bm v}_i$ and ${\bm v}_j$ are a pair of opposite vertices in a face.
As an example, a rigid cube has 8 nodes (24 DoF), 12 edges (12 edge length constraints) and 6 faces (6 diagonal length constraints). The remaining number of DoF is $24-12-6=6$ which corresponds to the three translational and the three rotational DoF.

Now, adding a link between two vertices ${\bm v}_i = (x_{3i-2}, x_{3i-1}, x_{3i})$ and ${\bm v}_j = (x_{3j-2}, x_{3j-1}, x_{3j})$ in two neighboring cubes imposes three \textit{link constraints}:
\begin{equation}
\left\{
\begin{aligned}
g_{\text{link}_x}({\bm v}_i,{\bm v}_j) &= x_{3i-2} - x_{3j-2} = 0,\\
g_{\text{link}_y}({\bm v}_i,{\bm v}_j) &=x_{3i-1} - x_{3j-1} = 0,\\
g_{\text{link}_z}({\bm v}_i,{\bm v}_j) &=x_{3i}- x_{3j} = 0.\\
\end{aligned}\right.
\end{equation}
We note that the decrease in DoF by adding a link can either be $0,1,2$, or $3$. If $n$ links are added to the $L \times M \times N$ prismatic assembly, there will be in total $18LMN+3n$ constraints ($18LMN$ length constraints, and $3n$ link constraints). To determine the infinitesimal DoF $d$ of the prismatic assembly, it is necessary to count the number of independent constraints. This can be done by the rigidity matrix rank computation~\cite{guest2006stiffness,davis2011algorithm}
\begin{equation}\label{dofcalculation}
d = 24LMN - \text{rank}(A),
\end{equation}
where $A$ is a rigidity matrix with the dimension $(18LMN+3n) \times 24LMN$, and $A_{ij} = \frac{\partial g_i}{\partial x_j}$ for all $i,j$ storing the partial derivatives of all above-mentioned constraints. Here the factor 24 stems from the fact that there are eight vertices for each solid cube, and for each vertex there are three coordinates.

%\subsubsection{The optimal lower bound of links and the hierarchical construction}
To determine the optimal lower bound of links needed to rigidify the assembly, we denote $\delta_{3D}(L,M,N)$ as the minimum number of links for rigidifying an $L \times M \times N$ prismatic assembly. Then we have
\begin{equation}
6LMN - 3\delta_{3D}(L,M,N) \leq 6.
\end{equation}
This implies that
\begin{equation}
\delta_{3D}(L,M,N) \geq \frac{6LMN-6}{3} = 2LMN-2.
\end{equation}
It is natural to ask whether the above lower bound is optimal (tight) for any combination of positive integers $L,M,N$. Denote a link pattern with exactly $2LMN-2$ links which can rigidify an $L \times M \times N$ prismatic assembly as a \textit{minimum rigidifying link pattern} (MRP) for $L \times M \times N$. Below, we devise a \textit{hierarchical construction} method for creating MRPs for infinitely many $L,M,N$. 

To illustrate the idea of the hierarchical construction, here we first consider the case where $L = M= N$ and simplify the notation $\delta_{3D}(L,M,N)$ as $\delta_{3D}(L)$. Suppose MRPs exist for $l_1 \times l_1 \times l_1$ and $l_2 \times l_2 \times l_2$, i.e. $\delta_{3D}(l_1) = 2l_1^3 -2$ and $\delta_{3D}(l_2) = 2l_2^3 - 2$. If we treat an $l_1 l_2 \times l_1 l_2 \times l_1 l_2$ prismatic assembly as $l_2 \times l_2 \times l_2$ large blocks with size $l_1 \times l_1 \times l_1$, we can rigidify each large block using an MRP for $l_1 \times l_1 \times l_1$ (which consists of exactly $\delta_{3D}(l_1)$ links) and then rigidify the entire structure using an MRP for $l_2 \times l_2 \times l_2$ (which consists of exactly $\delta_{3D}(l_2)$ links). Thus, the whole $l_1 l_2 \times l_1 l_2 \times l_1 l_2$ prismatic assembly is rigidified, with the total number of links
\begin{equation}
\begin{split}
l_2^3 \delta_{3D}(l_1)+\delta_{3D}(l_2)&=l_2^3 (2l_1^3-2) + (2l_2^3-2)\\
&=2(l_1l_2)^3-2.
\end{split}
\end{equation}
This suggests that the link pattern constructed this way is an MRP for $l_1 l_2 \times l_1 l_2 \times l_1 l_2$, i.e. $\delta_{3D}(l_1 l_2) = 2(l_1l_2)^3-2$. Using this idea of constructing larger MRPs via a hierarchical combination of smaller MRPs, we can prove that MRPs exist for all $L \times L \times L$ prismatic assembly with $L \geq 2$:

\begin{figure*}[t]
\centering
\includegraphics[width=0.7\textwidth]{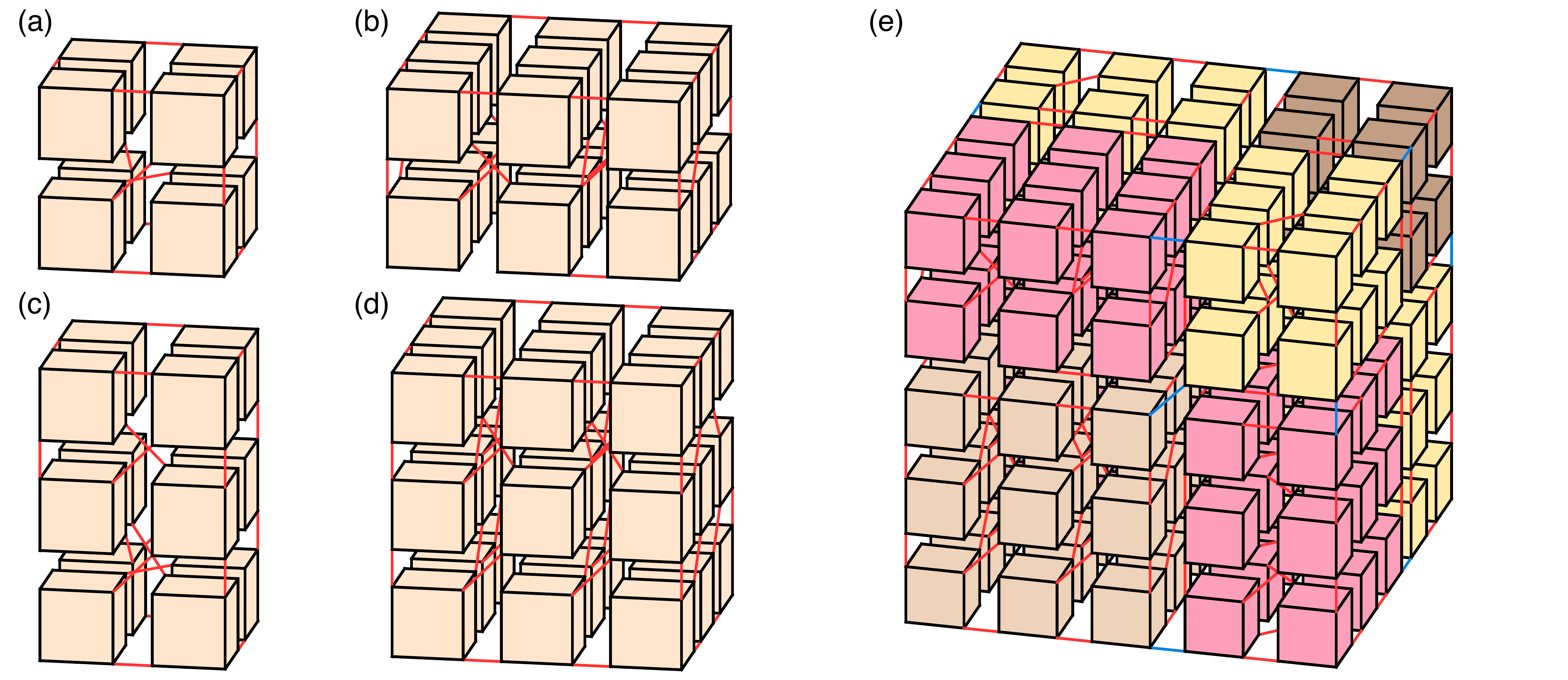}
\caption{{\bf Minimum rigidifying link patterns (MRPs) and the hierarchical construction protocol for prismatic assemblies.} (a)~An MRP with exactly $2\times 2^3-2 = 14$ links for a $2\times 2\times 2$ prismatic assembly. (b)~An MRP with exactly $2\times 2\times 3\times 3 -2 = 34$ links for a $2\times 3\times 3$ prismatic assembly. (c)~An MRP with exactly $2\times 2\times 2\times 3 -2 = 22$ links for a $2\times 2\times 3$ prismatic assembly. (d)~An MRP with exactly $2\times 3\times 3\times 3 -2 = 52$ links for a $3\times 3\times 3$ prismatic assembly. For all four examples, we have checked that DoF $=6$ using the rigidity matrix rank computation. (e) To construct a minimum rigidifying link pattern (MRP) for a $5\times 5 \times 5$ prismatic assembly, we treat the $5\times 5 \times 5$ cubes as eight large rectangular blocks with size $2\times 2 \times 2$, $2\times 2 \times 3$, $2\times 3 \times 3$, and $3\times 3 \times 3$ (each type is shown in a different color). We rigidify each block using an MRP in (a)-(d) (the red links), and then connect and rigidify the entire structure using an MRP for $2\times 2\times 2$ (the blue links), thereby obtaining an MRP for $5\times 5 \times 5$ (see text for details).}
\label{fig:F2}
\end{figure*}

\begin{theorem}\label{thm:rigidifying_3D}
For all positive integer $L \geq 2$, we have
\begin{equation}
\delta_{3D}(L) = 2L^3-2.
\end{equation}
\end{theorem}
{\it Proof.} We first explicitly construct MRPs for $L \times M \times N = 2\times 2 \times 2$, $3\times 3 \times 3$, $2\times 2 \times 3$, $2\times 3 \times 3$, each with exactly $2LMN-2$ links (Fig.~\ref{fig:F2}(a)-(d)), with the DoF of these assemblies verified computationally using Eq.~\eqref{dofcalculation}. The existence of such patterns shows that the statement is true for $L=2,3$. 

For $L\geq 4$, we prove the statement by induction. Suppose the statement is true for all positive integers less than $L$. Note that for $L\geq 4$, there always exists nonnegative integers $a,b$ with $a+b\geq 2$ such that $L = 2a + 3b$. To see this, we consider the following three cases:
\begin{enumerate}[(i)]
\item If $L \equiv 0 \text{ (mod } 3)$, we have $L = 2\times 0 + 3 \times \frac{L}{3}$.
\item If $L \equiv 1 \text{ (mod } 3)$, we have $L = 2\times 2 + 3 \times \frac{L-4}{3}$.
\item If $L \equiv 2 \text{ (mod } 3)$, we have $L = 2\times 1 + 3 \times \frac{L-2}{3}$.
\end{enumerate}

Now, we decompose the $L\times L\times L$ prismatic assembly into $(a+b) \times (a+b) \times (a+b)$ blocks with size $2\times 2 \times 2$, $2\times 2 \times 3$, $2\times 3 \times 3$, and $3\times 3 \times 3$ (Fig.~\ref{fig:F2}(e)). Since $2\leq a+b<L$, by the induction hypothesis, the number of links connecting these blocks is
\begin{equation}
\delta_{3D}(a+b) = 2(a+b)^3-2.
\end{equation}

Therefore, if we first rigidify each block by the corresponding MRP in Fig.~\ref{fig:F2}(a)-(d) and then rigidify the entire structure by an MRP for $(a+b) \times (a+b) \times (a+b)$, we obtain a rigidifying link pattern for the $L\times L\times L$ prismatic assembly, with the total number of links
\begin{equation}
\begin{split}
&a^3 \delta_{3D}(2) + b^3 \delta_{3D}(3) + 3a^2 b\delta_{3D}(2,2,3) \\
&+ 3ab^2 \delta_{3D}(2,3,3) + \delta_{3D}(a+b)\\
=&14 a^3 + 52 b^3 + 66a^2 b + 102 ab^2 + 2(a+b)^3-2\\
%=&16 a^3 + 54 b^3 + 72a^2 b + 108 ab^2 -2\\
=&2(2a+3b)^3 -2 = 2L^3-2.
\end{split}
\end{equation}
This implies that $\delta_{3D}(L) = 2L^3-2$. By induction, the statement is true for all $L \geq 2$. \hfill $\blacksquare$

Furthermore, we can explicitly construct MRPs for infinitely many $L,M,N$:
\begin{theorem}\label{thm:rigidifying_3D_LMN}
For infinitely many positive integers $L$, $M$, $N$ which are not all identical, we have
\begin{equation}
\delta_{3D}(L,M,N) = 2LMN-2.
\end{equation}
\end{theorem}
{\it Proof.} Take any set of nonnegative integers $a_l$, $b_l$, $a_m$, $b_m$, $a_n$, $b_n$ such that $a_l + b_l = a_m + b_m = a_n + b_n \geq 2$ and
\begin{equation}
\left\{
\begin{aligned}
L &= 2a_l + 3b_l,\\
M &= 2a_m + 3b_m,\\
N &= 2a_n + 3b_n
\end{aligned}\right.
\end{equation}
are not all identical (e.g. $(a_l, b_l, a_m, b_m, a_n, b_n) = (1,6,2,5,3,4)$, with $(L, M, N)=(20, 19, 18)$). Then, we can decompose an $L\times M \times N$ prismatic assembly into $(a_l+b_l) \times (a_m+b_m) \times (a_n+b_n)$ small blocks of size $2\times 2 \times 2$, $2\times 2 \times 3$, $2\times 3 \times 3$, and $3\times 3 \times 3$. Since $a_l + b_l = a_m + b_m = a_n + b_n$, following the proof of Theorem~\ref{thm:rigidifying_3D}, we rigidify each small block and then the entire structure using MRPs for different sizes. The total number of links of such a rigidifying link pattern for $L\times M \times N$ is
\begin{equation}
 \resizebox{0.95\hsize}{!}{$
\begin{split}
&a_l a_m a_n \delta_{3D}(2) + b_l b_m b_n \delta_{3D}(3) \\
&+ (a_l a_m b_n + a_la_n b_m + a_m a_n b_l) \delta_{3D}(2,2,3) \\&+ (a_l b_m b_n + a_m b_l b_n + a_n b_l b_m) \delta_{3D}(2,3,3) + \delta_{3D}(a_l+b_l)\\
=&14 a_l a_m a_n + 52 b_l b_m b_n + 22(a_l a_m b_n + a_la_n b_m + a_m a_n b_l) \\&+ 34 (a_l b_m b_n + a_m b_l b_n + a_n b_l b_m) \\&+ 2(a_l+b_l)(a_m+b_m)(a_n+b_n)-2\\
=&2(2a_l+3b_l)(2a_m+3b_m)(2a_n+3b_n) -2 =2LMN-2.
\end{split}
$}
\end{equation}
This implies that MRPs exist for $L\times M\times N$ and we have $\delta_{3D}(L,M,N) = 2LMN-2$. \hfill $\blacksquare$ 

We remark that the technique in the proof above can be used recursively for constructing more MRPs. For instance, as $2+0 = 1+1 = 0+2 = 2$, by considering $(a_l, b_l, a_m, b_m, a_n, b_n) = (2,0,1,1,0,2)$, we can construct an MRP for $4\times 5 \times 6$. Then, for any $a_l, b_l, a_m, b_m, a_n, b_n \geq 0$ with $a_l + b_l = 4, a_m + b_m = 5, a_n + b_n = 6$, we can use the same technique to construct an MRP for a $(2a_l+3b_l)\times (2a_m+3b_m)\times (2a_n+3b_n)$ prismatic assembly (see Section S1 in SI).

%\subsection{Minimum connecting link patterns (MCPs)}
Next, we consider the \textit{minimum connecting link patterns} (MCPs), i.e. link patterns with the minimum number of links that can connect all cubes in a prismatic assembly. Denote $\gamma_{3D}(L,M,N)$ as the minimum number of links needed for connecting an $L \times M \times N$ prismatic assembly. Since $\text{NCC} = LMN$ when there is no link, and each link reduces the NCC by at most one, the minimum number of links is $\gamma_{3D}(L,M,N) = LMN - 1$. To construct MCPs, one may make use of the hierarchical construction with the building blocks being four MCPs for $2\times 2 \times 2$, $2\times 2 \times 3$, $2\times 3 \times 3$, and $3\times 3 \times 3$, which can be easily constructed (see Section S2 in SI). Interestingly, $\delta_{3D}(L)$ and $\gamma_{3D}(L)$ are related by a simple formula:
\begin{equation}
\delta_{3D}(L) = 2L^3 - 2 = 2(L^3-1) = 2\gamma_{3D}(L).
\end{equation}
In other words, the minimum number of links needed for rigidifying any $L \times L \times L$ prismatic assembly is exactly twice of that for connecting it.

%\subsection{Simultaneous control of rigidity and connectivity using prescribed links}
Denote the DoF in a prismatic assembly by $d$ and the NCC by $c$. Using MRPs and MCPs, we can easily control the rigidity and connectivity of prismatic assembly and achieve different values of $d$ and $c$ simultaneously. 

For any $L \times M \times N$ prismatic assembly with an MRP, we have $d =6$ and $c = 1$. Further adding links to it will not change either $n$ or $c$, while removing any $k$ links from it will lead to an increase in $d$, making $d = 6+3k$. $c$ will remain unchanged until $\delta_{3D}(L,M,N) - k$ reaches a certain threshold. We may also obtain a prismatic assembly with $d = 6pqr$ and $c = pqr$, where $p|L$, $q|M$ and $r|N$. This is achieved by reversing the process of the hierarchical construction and remove links at the coarsest level from an MRP.

For any $L \times M \times N$ prismatic assembly with an MCP, it is clear that $c = 1$ and $d = 6LMN - 3 \gamma_{3D}(L,M,N) = 3LMN + 3$. Removing any $k$ links from it will lead to an increase in $c$ by $k$ and an increase in $d$ by $3k$, making $c = k+1$ and $d = 3LMN+3k+3$. We remark that the maximum internal DoF (DoF related to relative rotational motion among the assemblies rather than the rigid body motion) of an $L\times M \times N$ prismatic assembly is $3LMN - 3$, which is achieved if the link pattern is an MCP as MCPs connect all cubes and reduce the total translational and rotational DoF to the minimum (six).
 
\begin{figure}[t!]
\centering
\includegraphics[width=0.43\textwidth]{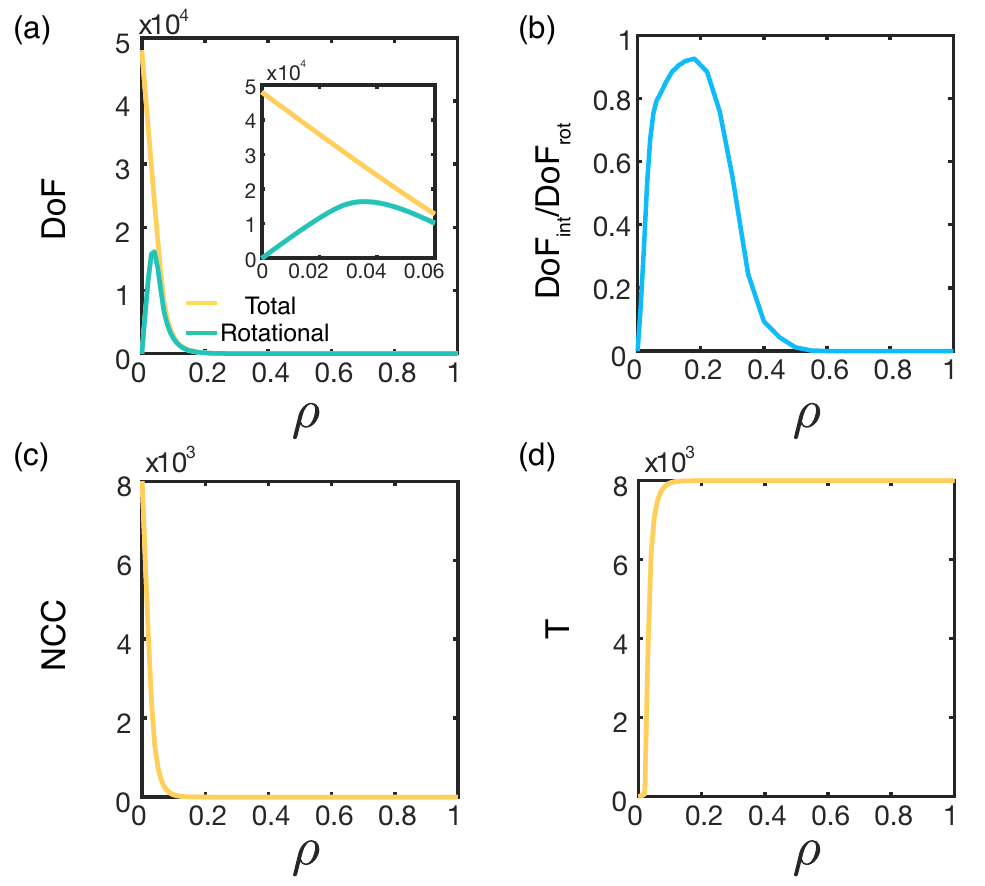}
\caption{{\bf Stochastic control of prismatic assemblies.} (a)~The total DoF (yellow) and the internal rotational DoF (green) with varying link density $\rho$, with a zoom-in of the behavior from $\rho = 0$ to 0.06 (inset). (b)~The ratio between the rotational DoF and the total DoF. (c)~The NCC with varying $\rho$. (d)~The size of the largest connected component ($T$) with varying $\rho$. The simulations are performed using a $20 \times 20 \times 20$ rectangular prismatic assembly.}%Data shown here are for an assembly of 8000 rectangular prisms similar to those shown in Fig. 2 }
\label{fig:F3}
\end{figure}

%\section{Stochastic control of prismatic assembly}
Having explored the deterministic control of rigidity and connectivity in a prismatic assembly, we now explore controlling these quantities by adding or removing links randomly, a process that we will see leads to percolation transitions in rigidity and connectivity. We note that in an $L\times L \times L$ prismatic assembly, the total number of possible links is $n_{\text{links}} =4(L-1)(7L^2-5L+1)$ (see Section S3 in SI). Denote $\rho \in [0,1]$ as the link density, i.e. the ratio of links randomly selected among all $n_{\text{links}}$ possible links in the prismatic assembly. We sample random links with different $\rho$ and study the DoF, NCC, and the size of the largest connected component of the resulting prismatic assembly. Fig.~\ref{fig:F3} shows the simulation results using a $20\times 20\times 20$ prismatic assembly, which consists of 64,000 nodes (192,000 coordinates). We observe that the total DoF decreases rapidly as $\rho$ increases, while the internal rotational DoF first increases and then decreases sharply~(Fig.~\ref{fig:F3}(a)). By increasing both the sampling frequency in between $\rho = 0$ and $\rho = 0.06$ and the number of repeats, we find that the peak of the internal DoF is at 0.036 (Fig.~\ref{fig:F3}(a) inset). 

In the deterministic case, the maximum internal DoF is achieved by MCPs at the link density
%it can be observed that the peak of the maximum internal DoF for prismatic assembly is attained at $\rho \approx 0.036$, which is much smaller than the result for planar prismatic assembly~\cite{chen2020deterministic}. To explain this, note that
\begin{equation}
 \resizebox{0.85\hsize}{!}{$\displaystyle
\frac{\gamma_{3D}(L)}{n_{\text{links}}}
= \frac{L^3-1}{4(L-1)(7L^2-5L+1)} 
\to \frac{1}{28} \approx 0.0357,
$}
\end{equation}
%Even if we only consider vertical or horizontal links, we still have $\frac{L^3-1}{12L^2(L-1)} \to \frac{1}{12} \approx 0.083$. 
which is very close to the peak density in the stochastic case. This is because when there are very few links, each newly added link is highly unlikely to be redundant, and thus most links reduce the total DoF by 3 and the NCC by 1, and increase the internal DoF by 2, until the assembly reaches the maximally floppy state. We note that the density at which the number of internal DoF reaches a maximum is much smaller than that in the planar analog $\frac{L^2-1}{4L(L-1)} \to \frac{1}{4} = 0.25$~\cite{chen2020deterministic} as there are many more possible links in the 3D case. Furthermore the number of rotational DoF is dominant among the total number of DoF, the ratio of which attains its maximum at $\rho \approx 0.15$ (Fig.~\ref{fig:F3}(b)). Both the range of dominance and its peak density are smaller than those in the 2D case, because the NCC decreases sharply in 3D (Fig.~\ref{fig:F3}(c)); the rigid body DoF reaches the minimum (six) quickly but some internal rotational DoF remain. The above maximum number of rotational DoF can also be understood in terms of the percolation transition in connectivity; indeed the size of the largest of connected component $T$ reaches $1/2$ at a similar link density (Fig.~\ref{fig:F3}(d)). A more detailed analysis of the finite size scaling confirms this (see Section S4 in SI). Our simulations suggest that we can easily obtain 3D assemblies with DoF, NCC or $T$ approximately equal to given target values by using random links within a rather small range of $\rho$. 

%\section{Discussion}
Our study of the topological control of prismatic assemblies provides novel strategies for achieving rigidity and connectivity via deterministic or stochastic cuts (links), thereby yielding new insights into the design of structural assemblies. We note that the metric constraints associated with infinitesimal rigidity for solid cubes can be naturally extended for any rectangular solids, and hence our results for the deterministic and stochastic control hold for general rectangular prismatic assemblies. It is also possible to extend our results to other space-filling prisms such as the triangular prisms (see Section S5 in SI). A natural next step is to explore rigidity and connectivity control of 3D assemblies formed by a tessellation of other polyhedra, including other space-filling polyhedra and their relatives such as the octet truss~\cite{fuller1961octet} or a combination of tetrahedra and octahedra~\cite{conway2011new}.  %While we have focused on the case of rectangular prismatic assemblies in our discussion, it is possible to extend our results to other space-filling prisms such as the triangular prisms (see Section S5 in SI). A natural next step is to explore the rigidity and connectivity control of 3D assemblies formed by a tessellation of other polyhedra such as the octet truss~\cite{fuller1961octet} or a combination of tetrahedra and octahedra~\cite{conway2011new}. 

%Our study complements these results in a few ways: the symmetry of the prismatic assembly naturally suggests the maximum coordination number at a vertex, and we start with a complete solid and consider the reverse problem of introducing cuts in it to determine the resulting topology and thence the connectivity and rigidity, rather than the inverse problem which is more common in percolation theory. 

{\bf Acknowledgments} This work was supported in part by the National Science Foundation Grant DMR 14-20570 (to L.M.) and DMREF 15-33985 (to L.M.).

\bibliographystyle{apsrev4-1_with_title}
\bibliography{kirigami3dcontrol_bib}

\end{document}